\documentclass{templateRevistaUPTCCiencias}
\usepackage{times,mathptmx}
\usepackage{lastpage}
\usepackage{fancyhdr}
\usepackage{float}
\usepackage{eucal}
\spanishdecimal{.}
\usepackage{cite}

\begin{document}
\fancypagestyle{plain}
{%
\fancyhf{}
\renewcommand{\headrulewidth}{0.5pt}
\renewcommand{\footrulewidth}{0pt}
}

\fancyhead{}
\fancyhead[CE]{\footnotesize{J. M. Ladino y E. Larra\~{n}aga}}
\fancyhead[CO]{\footnotesize{J. M. Ladino y E. Larra\~{n}aga}}

\fancyfoot{}
\fancyfoot[RO]{\footnotesize{\thepage}}
\fancyfoot[LE]{\footnotesize{\thepage}}
\renewcommand{\headrulewidth}{1pt}
\setlength{\columnsep}{6.5mm}

\title{Implementaci\'on computacional para el estudio de part\'iculas con esp\'in en agujeros negros rotantes modificados cu\'anticamente}
\author{
J. M. Ladino$^{1}$ y E. Larra\~{n}aga$^{2}$\\
\small{$^{1}$Universidad Nacional de Colombia. Sede Bogot\'a. Facultad de Ciencias. Observatorio Astron\'omico Nacional.\\ Bogot\'a, Colombia.} Correo electr\'onico:
\small{\texttt{jmladinom@unal.edu.co}}\\
\small{$^{2}$Universidad Nacional de Colombia. Sede Bogot\'a. Facultad de Ciencias. Observatorio Astron\'omico Nacional.\\ Bogot\'a, Colombia.} Correo electr\'onico:
\small{\texttt{ealarranaga@unal.edu.co}}
}
\small{\date{Octubre 17, 2022}}
\onecolumn
\maketitle
\begin{resumen}
Las trayectorias de part\'iculas de prueba (PP) con esp\'in que orbitan a un Agujero Negro (AN) dependen de la geometr\'ia del espacio-tiempo y de las caracter\'isticas f\'isicas intr\'insecas de la PP. En este trabajo se estudia el movimiento de PP con esp\'in sobre la \'orbita circular estable m\'as interna de un Agujero Negro Rotante (ANR) modificado cu\'anticamente. Se presenta una implementaci\'on computacional a trav\'es del programa Mathematica, la cual, puede calcular y graficar el radio, el momento angular y la energ\'ia en funci\'on del esp\'in de la PP y para diversos par\'ametros geom\'etricos del ANR, como bien lo son el esp\'in $a$ y la masa $M$ del ANR y los par\'ametros  cu\'anticos $\gamma$ y $\omega$. Adicionalmente, el programa calcula el esp\'in m\'aximo f\'isicamente posible para la PP.  Al final, se discuten algunos resultados dados por el programa y se muestra que estos son consistentes con los previamente reportados, tanto para los AN cl\'asicos de  Schwarzschild y Kerr como para los modificados cu\'anticamente.

\end{resumen}

\begin{abstract}
The trajectories of test particles with spin orbiting a black hole depend on the geometry of space-time and the intrinsic physical characteristics of the test particles. In this work the motion of test particles with spin on the innermost stable circular orbit of a quantum improved rotating black hole is studied. A computational implementation is presented through the Mathematica program, which can calculate and graph the radius, angular momentum and energy as a function of the spin of the test particles and for various geometric parameters of the rotating black hole, as they are the spin $a$ and the mass $M$ of the rotating black hole and the quantum parameters $\gamma$ and $\omega$. Additionally, the program calculates the maximum spin physically possible for the test particles. At the end, some results given by the program are discussed and it is shown that these are consistent with those previously reported, both for the classical black holes of Schwarzschild and Kerr and for the quantum improved ones.
\end{abstract}

\newpage
\twocolumn


\section{Introducci\'on}
Algunos objetos astrof\'isicos, como estrellas o Agujeros Negros (ANs) estelares, se encuentran orbitando alrededor de ANs supermasivos en los centros gal\'acticos. Es bien sabido que estos objetos tambi\'en se encuentran girando sobre s\'i mismos, por lo que estos pueden ser modelados como part\'iculas de prueba (PP) con un esp\'in cl\'asico. Las geod\'esicas que describen a la trayectoria de las PP que orbitan a un agujero negro han sido bastante estudiadas tanto en la teor\'ia de la Relatividad General como en diversas teor\'ias gravitacionales alternativas, demostrando que estas dependen de la geometr\'ia del espacio-tiempo y de las caracter\'isticas de la PP \cite{Zhang2018,Zhang2019,Zhang2020,Pugliese2013,Pugliese2011,  Suzuki1998,Jefremov2015, Yang2022, Conde2019, Frolov2003, Toshmatov2019,Abdujabbarov2010,   Larranaga2020, Abdujabbarov2011, Hackmann2010, Frolov2010,  Cariglia2012}. Por esta raz\'on, en los \'ultimos a\~{n}os numerosas investigaciones se han enfocado en analizar el efecto que tienen propiedades como la carga el\'ectrica o el esp\'in sobre las \'orbitas de las PP. El efecto del esp\'in cl\'asico de la PP est\'a descrito por las ecuaciones de movimiento de Mathisson-Papapetrou-Dixon (MPD) \cite{Mathison1937, Papapetrou1951, Dixon1970}. Un caso particular de gran inter\'es es el de una PP movi\'endose en la \'orbita circular estable m\'as interna (ISCO por su nombre en ingl\'es), de la cual se puede extraer informaci\'on \'util acerca de las propiedades del objeto central o de una posible estructura de acreci\'on y su espectro de radiaci\'on \cite{ Zhang2020}. 

En el a\~{n}o 2000, Bonanno y Reuter introdujeron una nueva soluci\'on tipo AN modificado mediante algunos argumentos de origen mec\'anico-cu\'antico \cite{Bonanno2000}. En esta soluci\'on se generaliza el espacio-tiempo est\'atico y esf\'ericamente sim\'etrico de Schwarzschild reemplazando la constante de gravitaci\'on universal de Newton $G_0$ por una funci\'on $G(r)$,\ dependiente de la distancia, e introduciendo dos par\'ametros cu\'anticos, $\gamma$ y $\omega$. La funci\'on $G(r)$ tiende a $G_0$, en el l\'imite $\omega \rightarrow 0$. Este formalismo  del grupo de renormalizaci\'on, adem\'as de ser uno de los m\'as famosos acercamientos hac\'ia la denominada gravedad cu\'antica, resuelve el problema de la singularidad en el origen de los ANs, i.e. esta es una  m\'etrica regular \cite{Bonanno2000}. Poco tiempo despu\'es, se present\'o en la literatura la correspondiente soluci\'on rotante (an\'aloga a la m\'etrica de Kerr, pero con la modificaci\'on cu\'antica) deducida a partir de diferentes m\'etodos \cite{Reuter2011, Bambi2013, Torres2017} y la cual tambi\'en representa a un AN regular. Previamente ya se han estudiado a las trayectorias de PP en los alrededores de estos ANs modificados cu\'anticamente \cite{Rayimbaev2020, Gao2021, Mandal2022,  Larranaga2022}. Adem\'as ya se han investigado otras aplicaciones y propiedades de los ANs modificados cu\'anticamente, como por ejemplo su sombra y \'angulo de reflexi\'on de la luz \cite{Kumar2020}.

En este trabajo se continua con la investigaci\'on recientemente publicada en la referencia \cite{Larranaga2022}. Inicialmente en la Secci\'on 2, se presenta al Agujero Negro Rotante (ANR) modificado cu\'anticamente deducido en  \cite{Torres2017}. Despu\'es, en la Secci\'on 3 se muestra, de manera simplificada, la deducci\'on de las ecuaciones de movimiento para una PP con esp\'in en el espacio-tiempo de un ANR modificado cu\'anticamente. El estudio del movimiento de las PP se restringe al plano ecuatorial de \'orbitas esp\'in-alineadas y esp\'in-antialineadas con el ANR para facilitar los c\'alculos. Luego, en la Secci\'on 4, se define al potencial efectivo para una PP con esp\'in moviendose en la ISCO de un ANR modificado cu\'anticamente. Adem\'as, se identifican los par\'ametros de la ISCO y se discute el movimiento superlum\'inico de la PP y su relaci\'on con el valor de esp\'in m\'aximo. En la Secci\'on 5, se presenta una implementaci\'on computacional a trav\'es de un cuaderno del programa Mathematica con el cual se puede calcular el esp\'in m\'aximo de la PP y estudiar los par\'ametros de la ISCO en funci\'on del esp\'in de la PP y para diversos par\'ametros geom\'etricos del ANR. En la secci\'on 6, se discuten varios resultados dados por el programa, analizando el comportamiento del esp\'in m\'aximo de la PP y mostrando c\'omo los resultados encontrados son consistentes, tanto con los previamente reportados para los AN de la Relatividad General (Schwarzschild y Kerr) como para los modificados cu\'anticamente. Por \'ultimo, en la secci\'on 7 se comparten algunas conclusiones.

\section{Agujero negro rotante modificado cu\'anticamente}
El espacio-tiempo de Schwarzschild modificado cu\'anticamente o del grupo de renormalizaci\'on \cite{Bonanno2000} es una soluci\'on tipo AN regular que se caracteriza por modificar $G_0$, la constante de gravitaci\'on universal de Newton, para convertirla en una funci\'on con una dependencia con respecto a un par\'ametro de escala de energ\'ia, $k$, el cual esta relacionado con la coordenada radial $r$. De modo que $G_0$ en este formalismo se transforma y se identifica con la funci\'on $G(r)$,  que en unidades de $c=1$, viene dada por
\begin{align}
G(r) = \frac{G_0 r^3}{r^3+  \omega G_0\left[r+\gamma G_0 M\right]}.\label{eq:runningNewtonconstant}
\end{align}
En esta anterior expresi\'on, $M$ es la masa del AN y $\omega$ y $\gamma$ son dos par\'ametros cu\'anticos constantes que provienen de la teor\'ia del grupo de renormalizaci\'on no perturbativa y de una correspondiente identificaci\'on de corte, respectivamente \cite{Bonanno2000,Torres2013}.  
Por otro lado, el caso rotante de la soluci\'on (\ref{eq:runningNewtonconstant}), se puede encontrar usando el bien conocido algoritmo de Newman-Janis \cite{Newman1965}. En esta ocasi\'on, el elemento de linea de un Agujero Negro Rotante (ANR) modificado cu\'anticamente en coordenadas de Boyer-Linquist generalizadas es  \cite{Torres2017}
\begin{align}
ds^{2}  = & -\left(1 - \frac{2MG(r)r}{\Sigma}\right)dt^{2} +  \frac{\Sigma}{\Delta} dr^{2}  \nonumber \\
 & +\Sigma d\theta^{2} -\frac{4MG(r)ar\sin^2 \theta }{\Sigma} dtd \phi \nonumber \\
 &+  \frac{\left( (r^2-a^2)^2 - a^2 \Delta \sin^2 \theta \right) \sin^2\theta}{\Sigma}  d\phi^{2},\label{eq:Metric}
\end{align}
con 
\begin{eqnarray}
\Delta & = & r^2+a^{2} - 2MG(r)r \\[1ex]
\Sigma & = & r^2 +a^{2}\cos^{2}\theta.
\end{eqnarray}
Los par\'ametros cu\'anticos se han calculado te\'oricamente realizando una comparaci\'on con la cuantificaci\'on perturbativa est\'andar de la gravedad de Einstein, obteniendo $\gamma=\frac{9}{2}$ y $\omega=\frac{167}{30 \pi}$ \cite{Bonanno2000,Torres2013}. No obstante, en este trabajo se tomar\'an a $\gamma$ y $\omega$  como par\'ametros libres, aprovechando el hecho de que las propiedades de la soluci\'on no dependen de sus valores precisos, siempre que estos sean estrictamente positivos\cite{Bonanno2000,Torres2013}.

El espacio-tiempo de la soluci\'on del ANR modificado cu\'anticamente, tiene tres casos particulares principales. La m\'etrica presentada en la ecuaci\'on (\ref{eq:Metric}), se reduce al AN est\'atico modificado cu\'anticamente cuando $a=0$, se convierte en un ANR de Kerr cuando $ \omega =0$ y reproduce a la soluci\'on de Schwarzschild cuando $a=0$ y $ \omega =0$ simult\'aneamente. En otras palabras, si $ \omega =0$, las correcciones cu\'anticas se anulan y se retorna a la gravedad cl\'asica descrita por las soluciones de la Relatividad General.  

Existen varias propiedades \'unicas para cada soluci\'on de AN, entre ellas, una de las m\'as importantes es el radio del horizonte de eventos. En este caso,  los respectivos horizontes modificados cu\'anticamente, se encuentran definidos por las ra\'ices de la expresi\'on $\Delta=0$, la cual corresponde a un polinomio de quinto grado y por tanto, no se pueden escribir anal\'iticamente. A\'un as\'i, expandiendo la funci\'on $G(r)\geq 0$ en una serie de potencias considerando una masa muy grande para el ANR, $M^2 \gg a^2$, se pueden encontrar los horizontes interior y exterior de manera aproximada como \cite{Torres2017}
\begin{equation}
r_{-}\approx\frac{1}{2} \sqrt{\gamma   \omega   G_0 + \sqrt{\gamma   \omega   G_0(8 a^2+\gamma   \omega  G_0)}}\label{eq:innerhorizon}
\end{equation}
y
\begin{equation}
r_+\approx G_0 M + \sqrt{G_0^2 M^2 - a^2} - \frac{(2+\gamma)\omega}{4M}.\label{eq:outerhorizon}
\end{equation}
Tomando $a=0$, los horizontes de las ecuaciones (\ref{eq:innerhorizon}) y (\ref{eq:outerhorizon}) son consistentes con los horizontes reportados para el AN de Schwarzschild modificado cu\'anticamente \cite{Torres2017}. Cuando $\omega = 0$, el horizonte de la ecuaci\'on \eqref{eq:outerhorizon} se reduce al valor correcto del horizonte exterior de la soluci\'on de Kerr, mientras que tomando $\omega = 0$ en la ecuaci\'on \eqref{eq:innerhorizon}, no se reproduce la  expresi\'on para el horizonte interno de la m\'etrica de Kerr. Esto ocurre como consecuencia del m\'etodo de aproximaci\'on \cite{Torres2017}.
\begin{figure*}[!htb]
    \centering\footnotesize
    \includegraphics[width=1\linewidth]{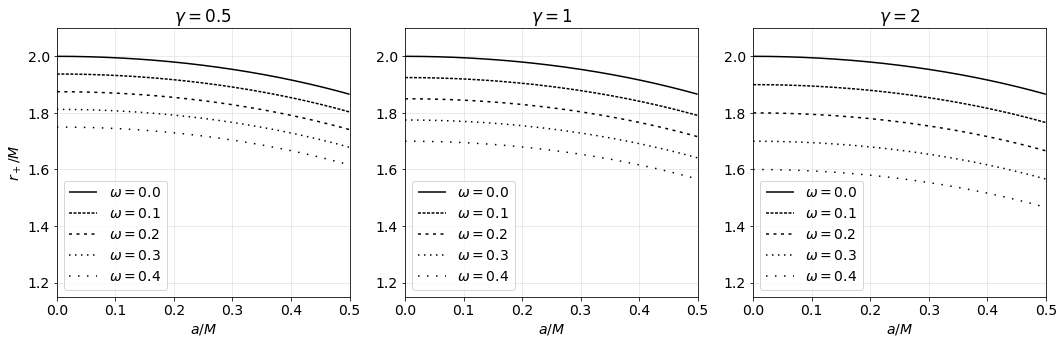}
    \caption{Radio del horizonte de eventos $r_+$ del ANR modificado cu\'anticamente en funci\'on del esp\'in $a$ del ANR y para diferentes valores de los par\'ametros $\omega$ y $\gamma$ (con $G_0 = M =1$).}
    \label{fig:HorizonRBH}
\end{figure*}


En la figura \ref{fig:HorizonRBH} se muestra el comportamiento del radio del horizonte externo del ANR modificado cu\'anticamente en funci\'on del esp\'in $a$ y para diferentes valores de los par\'ametros $\omega$ y $\gamma$. All\'i se evidencia que entre m\'as grandes sean las correcciones cu\'anticas dadas por $\omega$ y $\gamma$, el horizonte de eventos $r_+$ disminuir\'a cada vez m\'as. As\'i mismo, recordando que 
\begin{equation}
r_+=2G_0M\label{eq:Schwarzschildhorizon}
\end{equation}
en el AN de Schwarzschild y que 
\begin{equation}
r_{\pm}=  G_0M \pm \sqrt{G_0^2 M^2 - a^2}\label{eq:Kerrhorizon}
\end{equation}
para el ANR de Kerr, es claro que las curvas continuas de la figura \ref{fig:HorizonRBH}, que corresponden a $\omega=0$, describen correctamente a las soluciones de la Relatividad General. 

Es importante recordar que no cualquier valor de los par\'ametros geom\'etricos, $a$ y $M$,  y los par\'ametros  cu\'anticos, $\gamma$ y $\omega$, resultan describiendo una soluci\'on  tipo AN. Por ejemplo, en el caso cl\'asico del ANR de Kerr, existe una masa cr\'itica $M_c$ para la cu\'al, si $M<M_c$ no existen horizontes y por lo tanto la soluci\'on describir\'a una singularidad desnuda. El caso extremo del ANR de Kerr se obtiene cuando $M=M_c$, i.e. $a=G_0 M$, y corresponde a un AN en el cual los horizontes interno y externo coindicen en un \'unico horizonte con radio $r_{e}=G_0 M$. Este comportamiento para los ANR de Kerr se resume en la forma
\begin{equation}
\begin{cases}
\text{Si}\hspace{0.5cm}G_0 M = a& \hspace{0.5cm} \rightarrow \hspace{0.5cm}\text{es AN extremo}\\
\text{Si}\hspace{0.5cm}G_0 M > a& \hspace{0.5cm} \rightarrow \hspace{0.5cm}\text{es AN}\\
\text{Si}\hspace{0.5cm}G_0 M < a& \hspace{0.5cm} \rightarrow \hspace{0.5cm}\text{No es AN}.
\end{cases}
\label{eq:critickerr}
\end{equation}

Por lo tanto, para $G_0= M=1$, las soluciones de tipo ANR de Kerr son 
 validas si $0 \leq a \leq 1$.

En los ANs modificados cu\'anticamente no se tiene una soluci\'on anal\'itica con la informaci\'on completa de los horizontes y por ello debe identificarse num\'ericamente el valor cr\'itico $M_c$ y el correspondiente caso extremo. Haciendo $a=G_0 M$ y buscando num\'ericamente el valor de la ra\'iz real m\'as grande de $\Delta=0$ (que corresponder\'ia a $r_+$), este resulta ser un valor complejo para cualquier valor de los par\'ametros $\omega$ y $\gamma$, por lo que realmente no ser\'ia una soluci\'on de tipo AN. 

\begin{figure*}[!htb]
    \centering\footnotesize
    \includegraphics[width=1\linewidth]{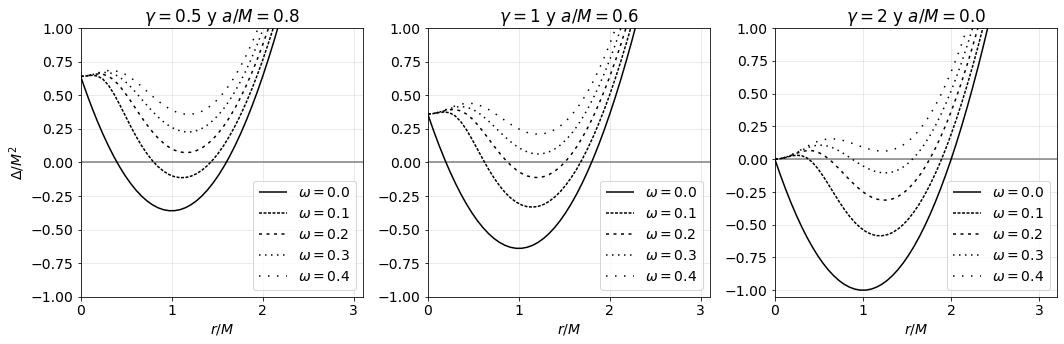}
    \caption{Funci\'on $\Delta$ para diferentes valores de los par\'ametros $a$, $\omega$ y $\gamma$ (con $G_0 = M =1$).}
    \label{fig:HorizonRBH2}
\end{figure*}

En la figura \ref{fig:HorizonRBH2} se ilustra el comportamiento de la funci\'on $\Delta$ para diferentes valores de los par\'ametros geom\'etricos  $a$, $\omega$ y $\gamma$. All\'i se evidencia  la existencia de una $M_c$ tanto para el caso est\'atico como para el rotante y por lo tanto, c\'omo no todos los valores de los par\'ametros geom\'etricos resultan en una soluci\'on del tipo AN. Por ejemplo, todas las curvas de la figura \ref{fig:HorizonRBH2} con $\omega=0.4$ representan \lq\lq singularidades\rq\rq  desnudas (aunque en este caso, debido a la modificiaci\'on cu\'antica, en realidad no hay  singularidades en el origen \cite{Torres2017}). Por otro lado, todas las curvas con $\omega=0.1$, con sus dos ra\'ices reales cerca del origen coordenado, si resultan en soluciones de tipo AN. 
 
\section{Ecuaciones de movimiento para part\'iculas con esp\'in}
Las ecuaciones que describen el movimiento de una PP con esp\'in cl\'asico en las cercan\'ias de un espacio-tiempo de ANR modificado cu\'anticamente son las ecuaciones de MPD. Bajo la aproximaci\'on \lq\lq polo-dipolo\rq\rq estas toman la forma \cite{Mathison1937, Papapetrou1951, Dixon1970}
\begin{align}
\label{eq:eom1}
\frac{DP^\mu }{D\tau} &= -\frac{1}{2} R^\mu _{ \ \nu \rho \sigma} u^\nu S^{\rho \sigma} \\[2ex]
\label{eq:eom2}
\frac{DS^{\mu \nu} }{D\tau} &= P^\mu u^\nu- P^\nu u^\mu,
\end{align}
con $u^\mu = \frac{dx^\mu}{d\tau}$ la 4-velocidad de la PP y $P^\mu$ su correspondiente 4-momentum, $ R^\mu _{ \ \nu \rho \sigma}$ es el tensor de  Riemann y $S^{\mu \nu} = - S^{\nu \mu}$ es el tensor de esp\'in, con el cual se define el esp\'in de la PP como
\begin{equation}
s^2 = \frac{1}{2} S^{\mu \nu} S_{\mu \nu}.
\label{eq:spin}
\end{equation}
 
 Ahora, para cerrar el sistema de ecuaciones, en esta ocasi\'on se incluir\'a a la condici\'on suplementaria de esp\'in de Tulczyjew,
\begin{equation}
P_\mu S^{\mu \nu} = 0,
\label{eq:Tulczyjew}
\end{equation}
y la normalizaci\'on del vector momentum
\begin{equation}
P_\mu P^\mu = - m^2,
\label{eq:momentum}
\end{equation}
donde $m$ es la masa propia de la PP \cite{Lukes2014,Zhang2018}. Cabe resaltar que existen otras condiciones suplementarias de esp\'in. Cada una de ellas fija de forma diferente un centro de referencia y por lo tanto cada una describe una evoluci\'on diferente de las ecuaciones MPD \cite{Lukes2014}. \\

El espacio-tiempo en el entorno del ANR mejorado cu\'anticamente admite dos vectores de Killing que est\'an vinculados con dos cantidades conservadas principales del sistema. Hay un vector de Killing como-de-tiempo $\xi^{\mu} =\frac{\partial}{\partial t}$ relacionado con la conservaci\'on de la energ\'ia $E=-C_{\xi}$ y otro como-de-espacio $ \varphi^{\mu} =  \frac{\partial}{\partial \phi}$ relacionado con la conservaci\'on del momentum angular total $J=C_{\varphi}$ \cite{Zhang2020,Saijo1998}.  Estas cantidades conservadas de la PP con esp\'in se relacionan con su respectivo vector de Killing $k$ mediante 
\begin{equation}
C_{k} = P^{\mu}k_\mu + \dfrac{1}{2}S^{\mu \nu} \nabla_{\mu} k_\nu.
\label{eq:conserved}
\end{equation}

Ahora bien, con el objetivo de simplificar el problema se van a considerar de ahora en adelante las siguientes restricciones:
\begin{itemize}
    \item La PP se mover\'a sobre una \'orbita circular
    \begin{equation}
        \dot{r}=0.
        \label{eq:res1}
    \end{equation}
      \item La PP tendr\'a una trayector\'ia sobre el plano ecuatorial del ANR
    \begin{equation}
        \theta = \frac{\pi}{2}.
        \label{eq:res2}
    \end{equation}
      \item La PP podr\'a tener dos posibles situaciones: estar en una \'orbita esp\'in-alineada ($s>0$) o esp\'in-antialineada ($s<0$) con respecto al plano ecuatorial del ANR
    \begin{equation}
        s^{\mu}=(0,0,-s,0).
        \label{eq:res3}
    \end{equation}
    En otras palabras, cuando la PP tiene una \'orbita esp\'in-alineada, su momentum angular de esp\'in ser\'a paralelo al momentum angular de esp\'in del ANR, mientras que con una \'orbita esp\'in-antialineada, su momentum angular de esp\'in ser\'a antiparalelo al momentum angular de esp\'in del ANR.
\end{itemize}
De modo que, teniendo en cuenta las anteriores restricciones y usando el formalismo de las tetradas \cite{Larranaga2022}, es posible encontrar mediante las ecuaciones \eqref{eq:momentum} y \eqref{eq:conserved} las componentes del 4-momentum,
\begin{equation}
\begin{cases}
P^0 = & \frac{1}{\Delta Z} \left[ (r^2 + a^2)X + ar \Delta Y \right]  \\
P^1 = & \pm \frac{\sqrt{\mathcal{R}}}{rZ} \\ 
P^2 = & 0  \\
P^3 = & \frac{1}{\Delta Z} \left[ aX + r \Delta Y\right]
\end{cases}
\end{equation}
con
\begin{equation}
\begin{aligned}
X = &\left[ r^3 + a^2r + as \left( r + M(G(r) - r G'(r))\right) \right] E \\ 
& - \left[ar + Ms(G(r) - r G'(r))  \right] J \\[2ex]
Y = &  J - (a+s)E  \\
Z =& r^3 - Ms^2 \left[ G(r) - r G'(r)\right] \\
\mathcal{R} = &r^2 X^2 - \Delta \left( r^4 Y^2 + m^2 Z^2 \right),
\end{aligned}
\label{eq:AuxFunctions}
\end{equation}
donde $G'(r)$ representa la derivada de $G(r)$ con respecto a la coordenada $r$. Luego, nuevamente bajo las restricciones de las ecuaciones \eqref{eq:res2} y \eqref{eq:res3} junto con las ecuaciones \eqref{eq:spin}, \eqref{eq:Tulczyjew} y \eqref{eq:momentum}, las componentes no nulas del tensor de esp\'in toman la forma
\begin{equation}
\begin{cases}
S^{01} = &-S^{01} = \dfrac{sP_{3}}{mr} \\
S^{03} = &-S^{30} = - \dfrac{sP_{1}}{mr} \\
S^{31} = &-S^{13} = \dfrac{sP_{0}}{mr}.
\end{cases}
\label{eq:spincomponents2}
\end{equation}

Remplazando las componentes no nulas del 4-momentum, del tensor de esp\'in y del tensor de Riemann en las ecuaciones \eqref{eq:eom1} y \eqref{eq:eom2}, se obtienen las siguientes ecuaciones para las velocidades \cite{Larranaga2022}
\begin{align}
\dot{r} = &P^1\left[ 1 + \frac{s^2}{m^2r^2}g_{11} R_{3003} \right] \notag \\
& \times \left[ P^0  + \frac{s^2}{m^2r^2} \left(R_{3113}P_0 + R_{3101}P_3\right) - \frac{s}{mr^3}P_3 \right]^{-1},
	\label{eq:radialvelocity}\\
\dot{\phi} = &\left[P^3 + \frac{s^2}{m^2 r^2} \left( R_{1001}P_3 + R_{1013}P_0 \right) - \frac{sP_1}{m r^2} \dot{r} \right] \notag  \\
& \times \left[ P^0 - \frac{s^2}{m^2 r^2} \left(R_{1301}P_3 + R_{1313}P_0 \right) \right]^{-1}.
	\label{eq:angularvelocity}
\end{align}

La deducci\'on detallada de estas anteriores expresiones junto con la forma expl\'icita de las componentes no nulas de $ R^\mu _{ \ \nu \rho \sigma}$ se pueden consultar en la referencia  \cite{Larranaga2022}. En el resultado de la ecuaci\'on \eqref{eq:radialvelocity}, claramente se evidencia que la velocidad radial $\dot{r}$ es paralela a $P^1$, la componente radial del 4-momentum. De esta manera, la restricci\'on de \'orbitas circulares de la ecuaci\'on  \eqref{eq:res1}, es equivalente a la condici\'on $P^1=0$, de la cual es posible identificar al potencial efectivo del sistema.

\section{\'Orbita ISCO y movimiento superlum\'inico} 
Introduciendo las cantidades $e = \frac{E}{m}$ y $j=\frac{J}{m} = \frac{\ell+s}{m}$ , energ\'ia y momentum angular total para la PP con esp\'in, siendo $\ell$ el momentum angular orbital, se puede reescribir la condici\'on de \'orbitas circulares como \cite{Larranaga2022}
\begin{align}
(P^1)^{2} = 0= &\frac{m^2}{r^2 Z^2} \left(A e^{2} + B e + C\right)\notag \\
	= &\frac{m^2 A}{r^2 Z^2} \left(e -\frac{-B+\sqrt{B^{2}-4AC}}{2A}\right) \notag \\
	&\times \left(e+\frac{B+\sqrt{B^{2}-4AC}}{2A}\right)
	\label{eq:squaremomentum}
\end{align}
donde 
\begin{align}
A =& r^2 \left[ K_1^2 - \Delta r^2 (a+s)^2 \right] \\
B =& 2 r^2  j \left[ K_1 K_2 - \Delta r^2 (a+s) \right] \\
C =& r^2 j^2 \left[ K_2^2 - \Delta r^2 \right] - \Delta Z^2\\
K_1 =& r^3 + a^2 r + as \left[ r + M(G(r) - r G'(r))\right]\\
K_2 =& - \left[ ar + Ms(G(r) - r G'(r) )\right].
\end{align}

Ahora bien, es posible definir al potencial efectivo de la PP con esp\'in, a partir de la ra\'iz positiva (porque el movimiento debe estar dirigido hacia el futuro \cite{Saijo1998}) de la ecuaci\'on \eqref{eq:squaremomentum}, as\'i
\begin{equation}
V_{\textrm{eff}} (r) =  \frac{-B + \sqrt{B^2 - 4 A C}}{2 A}.
\label{eq:V-effective}
\end{equation}
Luego, la \'orbita ISCO esta definida por las condiciones
\begin{align}
\frac{dV_{\textrm{eff}}}{dr} = 0
\label{eq:V-effectivederivate1}
\end{align}
y
\begin{align}
 \frac{d^2 V_{\textrm{eff}}}{dr^2} = 0 .
\label{eq:V-effectivederivate2}
\end{align}
El prop\'osito de este trabajo es analizar el comportamiento de los par\'ametros de la \'orbita ISCO que describe la PP con esp\'in. Estos par\'ametros son el radio $r_{ISCO}$, la energ\'ia por unidad de masa $e_{ISCO}$ y el momentum angular orbital por unidad de masa $\ell_{ISCO}$ de la \'orbita ISCO.

Los par\'ametros de la ISCO para PP sin esp\'in alrededor de un AN de Schwarzschild son \cite{Zhang2018}
\begin{align}
\begin{cases}
r_{ISCO} =& 6M \\  \ell_{ISCO} =&2\sqrt{3}M \approx 3.4641M \\   e_{ISCO}=&\sqrt{\frac{8}{9}}\approx 0.9428.
\label{eq:Schwarzschildsinspin}
\end{cases}
\end{align}

Por otro lado, debido al efecto de arrastre del ANR de Kerr, los par\'ametros de las ISCO son diferentes dependiendo de si la \'orbita es corrotante (momentum angular \'orbital paralelo al momentum angular de esp\'in del ANR) o contrarrotante con respecto a la rotaci\'on del ANR. Para PP sin esp\'in corrotantes y contrarotantes con el ANR de Kerr extremo ($a=M)$, los par\'ametros de la ISCO son respectivamente \cite{Zhang2018}
\begin{align}
\begin{cases}
r_{ISCO} =& M \\  \ell_{ISCO} =&\frac{2}{\sqrt{3}}M \approx 1.1547M \\   e_{ISCO}=&\frac{1}{\sqrt{3}}\approx 0.5774
\label{eq:kerrsinspin1}
\end{cases}
\end{align}
y
\begin{align}
\begin{cases}
r_{ISCO} =& 9M \\  \ell_{ISCO} =&-\frac{22}{3\sqrt{3}}M \approx -4.2339M \\   e_{ISCO}=&\sqrt{\frac{8}{9}}\approx 0.9622.
\label{eq:kerrsinspin2}
\end{cases}
\end{align}

De manera general, para PP sin esp\'in corrotantes y contrarotantes con el ANR de Kerr con $0\leq a \leq 1$, los intervalos posibles para los par\'ametros de la ISCO son respectivamente 
\cite{Zhang2019}
\begin{align}
\begin{cases}
\hspace{0.45cm} M \leq & r_{ISCO} \hspace{0.2cm}\leq  6M \\ \frac{2} {\sqrt{3} }M \leq & \ell_{ISCO} \hspace{0.2cm}\leq 
 2\sqrt{3}M \\ 
 \hspace{0.35cm} \frac{1}{\sqrt{3}}  \leq & e_{ISCO} \hspace{0.2cm}\leq \sqrt{\frac{8}{9}}
\label{eq:kerrsinspin3}
\end{cases}
\end{align}
y
\begin{align}
\begin{cases}
\hspace{0.75cm} 6M \leq & r_{ISCO} \hspace{0.2cm} \leq  9M \\ -\frac{22}{3\sqrt{3}}M \leq & \ell_{ISCO} \hspace{0.2cm} \leq 
 -2\sqrt{3}M \\ 
 \hspace{0.65cm} \sqrt{\frac{8}{9}}  \leq & e_{ISCO} \hspace{0.2cm} \leq \frac{5}{3\sqrt{3}}.
\label{eq:kerrsinspin4}
\end{cases}
\end{align}

En los an\'alisis de las siguientes secciones se considerar\'a \'unicamente el caso en que la PP se encuentra en una \'orbita corrotante con respecto a la rotaci\'on del ANR.

Ahora bien, los par\'ametros $r_{ISCO}$ y $\ell_{ISCO}$ para PP con esp\'in se pueden encontrar num\'ericamente usando las ecuaciones \eqref{eq:V-effectivederivate1} y \eqref{eq:V-effectivederivate2} y con ellos el par\'ametro $e_{ISCO}$  mediante la ecuaci\'on \eqref{eq:V-effective}.

Es importante notar que no cualquier valor del esp\'in $s$ de la PP describe una trayectoria f\'isicamente posible. Para ciertos valores de $s$, la 4-velocidad de la PP tendr\'ia una trayector\'ia como de espacio \cite{Zhang2020,Zhang2018}. Por lo tanto,  se impondr\'a sobre la 4-velocidad la condici\'on
\begin{equation}
    \frac{u^2}{(u^0)^2} = g_{00} + g_{11}\dot{r}^2+ 2 g_{03} \dot{\phi}  + g_{33} \dot{\phi}^2  < 0.
    \label{eq:superluminal}
\end{equation}

Esta condici\'on brinda un valor de esp\'in m\'aximo, $s_{max}$, por encima del cual se tendr\'a un movimiento superluminal (la PP tendr\'ia una velocidad mayor a la de la luz). Por ejemplo, el valor del esp\'in m\'aximo reportado en la literatura \cite{Zhang2018} para una PP en el espacio-tiempo de Schwarzschild es de $s_{max}=1.6510M$,  y con este, los par\'ametros de la ISCO son  
\begin{align}
\begin{cases}
r_{ISCO} \approx& 2.5308M \\  \ell_{ISCO}  \approx &1.3249M \\   e_{ISCO} \approx& 0.7896.
\label{eq:Schsinspin2}
\end{cases}
\end{align}

\section{Implementaci\'on computacional}

En el repositorio \cite{gitrepo} se incluye un cuaderno del programa Mathematica que calcula el radio, la energ\'ia y el momento angular para una PP con esp\'in movi\'endose alrededor de un ANR modificado cu\'anticamente. 

Para utilizar este c\'odigo, se debe ingresar los par\'ametros que determinan a la m\'etrica $(M,G_0,a,\omega, \gamma)$ y los valores iniciales de radio y momento angular $(r_{ISCO}^0, \ell_{ISCO}^0)$ para realizar la b\'usqueda de los par\'ametros de la ISCO. Tambi\'en es posible seleccionar si se desea exportar los datos obtenidos como un archivo. Despu\'es de ingresar esta informaci\'on, simplemente se debe evaluar todo el cuaderno y el resultado principal corresponde a una gr\'afica con los valores de radio, momento angular y energ\'ia para la ISCO de part\'iculas con spin entre $-2$ y $2$ (intervalo que se puede ajustar en el programa). En esta gr\'afica tambi\'en se indica el valor del esp\'in m\'aximo $s_{max}$  permitido para obtener una velocidad como-de-tiempo.

\section{An\'alisis y resultados} 

\begin{figure*}[!htb]
    \centering\footnotesize
    \includegraphics[width=1\linewidth]{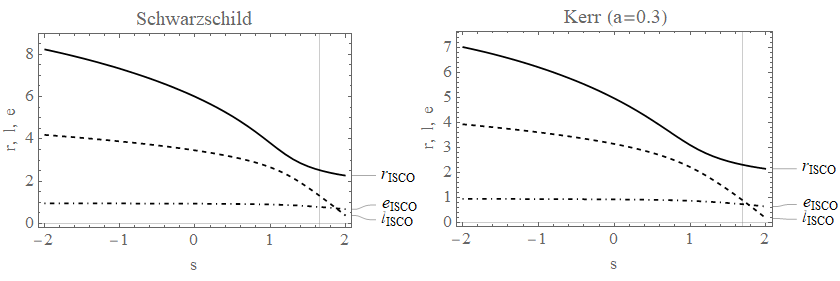}
    \caption{Gr\'aficas obtenidas por el programa de Mathematica de los par\'ametros de la ISCO en funci\'on del esp\'in $s$ de la PP (con $G_0 = M =1$). La gr\'afica de la izquierda corresponde al AN de Schwarzschild (con $a=\omega=\gamma=0$).  La gr\'afica de la derecha corresponde al ANR de Kerr (con $a=0.3$ y $\omega=\gamma=0$). La linea vertical gris muestra el valor del $s_{max}$.}
    \label{fig:COCOAmath1}
\end{figure*}
\begin{figure*}[!htb]
    \centering\footnotesize
    \includegraphics[width=1\linewidth]{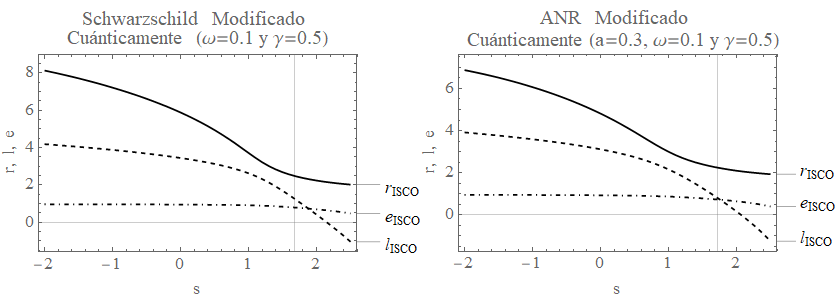}
    \caption{Gr\'aficas obtenidas por el programa de Mathematica de los par\'ametros de la ISCO en funci\'on del esp\'in $s$ de la PP (con $G_0 = M =1$, $\omega=0.1$ y $\gamma=0.5$). A la izquierda para el AN de Schwarzschild modificado cu\'anticamente (con $a=0$) y a la derecha para el ANR modificado cu\'anticamente (con $a=0.3$). La linea vertical gris muestra el valor del $s_{max}$.}
    \label{fig:COCOAmath3}
\end{figure*}
\begin{figure*}[!htb]
    \centering\footnotesize
    \includegraphics[width=1\linewidth]{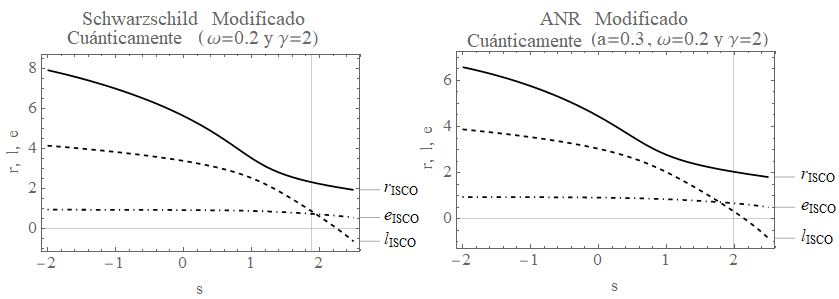}
    \caption{Gr\'aficas obtenidas por el programa de Mathematica de los par\'ametros de la ISCO en funci\'on del esp\'in $s$ de la PP (con $G_0 = M =1$, $\omega=0.2$ y $\gamma=2$). A la izquierda para el AN de Schwarzschild modificado cu\'anticamente (con $a=0$) y a la derecha para el ANR modificado cu\'anticamente (con $a=0.3$). La linea vertical gris muestra el valor del $s_{max}$.}
    \label{fig:COCOAmath2}
\end{figure*}

En las figuras \ref{fig:COCOAmath1}, \ref{fig:COCOAmath3}  y \ref{fig:COCOAmath2}  se ilustran las gr\'aficas obtenidas por el programa de Mathematica correspondientes al comportamiento de los par\'ametros de la ISCO en funci\'on del esp\'in de la PP. Los resultados mostrados en la figura \ref{fig:COCOAmath1} corresponden a los de a una PP movi\'endose en los espacio-tiempos de la Relatividad General:  la gr\'afica de la izquierda es para el AN de Schwarzschild mientras que la de la derecha es para el ANR de Kerr (con $a=0.3$). Las lineas verticales de color gris muestran el valor del esp\'in m\'aximo $s_{max}$ permitido por la condici\'on de una velocidad como de tiempo de la ecuaci\'on \eqref{eq:superluminal}.  Las figuras \ref{fig:COCOAmath3}  y \ref{fig:COCOAmath2}  revelan el comportamiento de los par\'ametros de la ISCO para una PP con esp\'in $s$ movi\'endose en los espacio-tiempos de los ANs modificados cu\'anticamente.

Los resultados dados en las figuras \ref{fig:COCOAmath1}, \ref{fig:COCOAmath3}  y \ref{fig:COCOAmath2} corroboran los previamente obtenidos en las referencias \cite{Larranaga2022,Zhang2018,Jefremov2015}. All\'i se muestra que los par\'ametros de la ISCO para los AN cl\'asicos ($ \omega =0$), tanto para el de Schwarzschild ($a=0$) como para el de Kerr ($a \neq 0$), disminuyen con un aumento del esp\'in $s$. 

Al considerar los resultados para el AN est\'atico modificado cu\'anticamente ($ \omega \neq 0$ y $a=0$) y del ANR modificado cu\'anticamente, los par\'ametros de la ISCO tambi\'en disminuyen con un aumento del esp\'in $s$ o con un aumento de los par\'ametros $ \omega $ y $\gamma$. Adem\'as, se evidencia que al introducir el par\'ametro de esp\'in de los ANRs $a\neq 0$ los par\'ametros de la ISCO son m\'as peque\~{n}os que los de los casos de los ANs no rotantes. 

Por otro lado, analizando los posibles valores que puede alcanzar el esp\'in m\'aximo $s_{max}$, en las figuras \ref{fig:COCOAmath3}  y \ref{fig:COCOAmath2}, los resultados muestran que para mayores valores de los par\'ametros cu\'anticos $\omega $ y $\gamma$ se obtienen mayores valores del esp\'in m\'aximo $s_{max}$ permitido. No obstante, aqu\'i no se ha identificado el efecto del par\'ametro de esp\'in $a$ de los ANRs sobre el valor del $s_{max}$.

\begin{figure*}[!htb]
    \centering\footnotesize
    \includegraphics[width=1\linewidth]{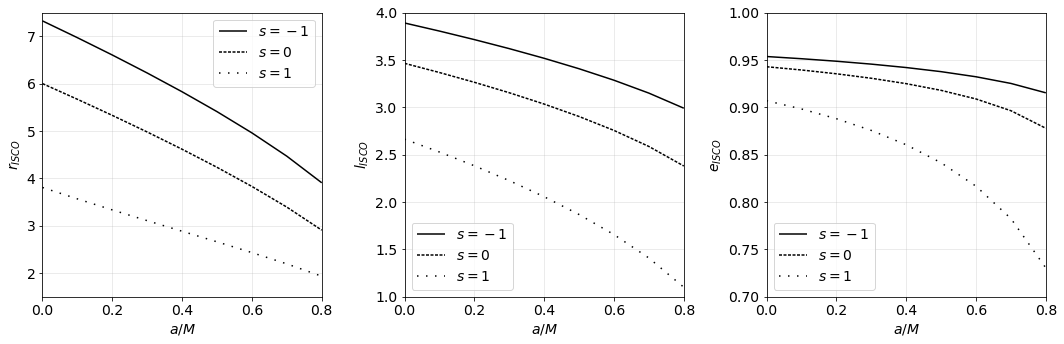}
    \caption{Comportamiento de los par\'ametros de la ISCO en funci\'on del esp\'in $a$ del ANR de Kerr y para diferentes valores del esp\'in de la PP (con $G_0 = M =1$).}
    \label{fig:RBH3}
\end{figure*}

\begin{figure*}[!htb]
    \centering\footnotesize
    \includegraphics[width=1\linewidth]{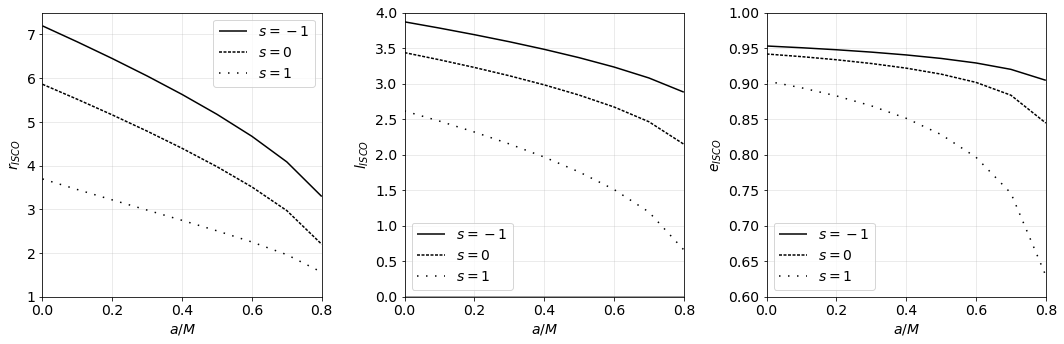}
    \caption{Comportamiento de los par\'ametros de la ISCO en funci\'on del esp\'in $a$ del ANR modificado cu\'anticamente y para diferentes valores del esp\'in de la PP  (con $\omega=0.1$, $\gamma=1$ y $G_0 = M =1$).}
    \label{fig:RBH4}
\end{figure*}

En las tablas \ref{tabla1}, \ref{tabla2}, \ref{tabla3}, \ref{tabla4}, \ref{tabla5} y  \ref{tabla6}, se muestran los resultados de los valores del $s_{max}$ obtenido por el programa de Mathematica para una PP en el espacio-tiempo de los ANs de la Relatividad General y los ANs modificados cu\'anticamente.
\begin{table}[H]
	\caption{\small{Valores del $s_{max}$ obtenido por el programa de Mathematica para una PP en el espacio-tiempo de los ANs de la Relatividad General (con $G_0=1$ y $\gamma=\omega=0$).}\label{tabla1}}
	\centering {\small
		\begin{tabular}{ccr}\hline
			$a/M$    &
			Tipo de AN &
			\multicolumn{1}{c}{$s_{max}/M$}\\ \hline
			0    & Schwarzschild  &  $1.6518$\\
			0.01    & Kerr  &  $ 1.6542$\\
			0.05    & Kerr  &  $1.6631$\\
			0.1    & Kerr  &  $ 1.6725$\\
			0.2    & Kerr  &  $1.6845$\\
			0.3    & Kerr  &  $1.6865$\\
			0.4    & Kerr  &  $ 1.6770$\\
			0.5    & Kerr  &  $1.6542$\\
			0.6    & Kerr  &  $1.6154$\\
			0.7    & Kerr  &  $ 1.5564$\\
            0.8    & Kerr  &  $ 1.4697$\\ \hline
   
		\end{tabular}}
	\end{table}
	
\begin{table}[H]
	\caption{\small{Valores del $s_{max}$ obtenido por el programa de Mathematica para una PP en el espacio-tiempo de Schwarzschild modificado cu\'anticamente (con $G_0=1$ y $a=0$).}\label{tabla2}}
	\centering {\small
		\begin{tabular}{cccc}\hline
			$\hspace{1cm}$ & $\hspace{1cm}$ & $s_{max}/M$ & $\hspace{1cm}$\\ \hline 
   $\omega$  &
		$\gamma=0.5$ & $\gamma=1$  & $\gamma=2$ \\ \hline
			$0.1$    & $ 1.6829$  &  $1.6994$ &  $1.7365$\\
			$0.2$    & $ 1.7245$  &  $1.7691$ &  $1.8880$\\
			$0.3$    & $ 1.7826$  &  $1.8796$ &  $2.2208$\\
   		$0.4$    & $ 1.8687$  &  $2.0767$ &  No es AN\\
     	$0.5$    & $2.0072$ &  $2.4868$ &  No es AN\\
       	$0.6$    & $2.2566$  &  No es AN &  No es AN\\
        	$0.7$    & No es AN  &  No es AN &  No es AN\\
\hline
		\end{tabular}}
	\end{table}

 \begin{table}[H]
	\caption{\small{Valores del $s_{max}$ obtenido por el programa de Mathematica para una PP en el espacio-tiempo de un ANR modificado cu\'anticamente (con $G_0=1$ y $\omega=0.1$).}  \label{tabla3}}
	\centering {\small
		\begin{tabular}{cccc}\hline
			$\hspace{1cm}$ & $\hspace{1cm}$ & $s_{max}/M$ & $\hspace{1cm}$\\ \hline 
   $a/M$  &
		$\gamma=0.5$ & $\gamma=1$  & $\gamma=2$ \\ \hline
			$0.01$    & $ 1.6854$  &  $1.7020$ &  $1.7392$\\
			$0.05$    & $ 1.6945$  &  $1.7114$ &  $1.7492$\\
			$0.1$    & $ 1.7041$  &  $1.7212$ &  $1.7598$\\
			$0.2$    & $ 1.7164$  &  $1.7341$ &  $1.7745$\\
   		$0.3$    & $ 1.7183$  &  $1.7366$ &  $1.7794$\\
			$0.4$    & $ 1.7082$  &  $1.7274$ &  $1.7731$\\
   		$0.5$    & $ 1.6839$  &  $1.7042$ &  $1.7545$\\
			$0.6$    & $ 1.6422$  &  $1.6639$ &  $1.7222$\\
   		$0.7$    & $ 1.5774$  &  $1.6014$ &  $1.6802$\\
      	$0.8$    & $ 1.4764$  &  $1.5083$ &  No es AN\\
            $0.9$    & No es AN  &  No es AN &  No es AN\\
\hline
		\end{tabular}}
	\end{table}

Los resultados dados en la tabla \ref{tabla1} muestran el valor del $s_{max}$ para diferentes valores posibles del par\'ametro de esp\'in $a$ del ANR de Kerr.
All\'i se evidencia que el efecto de $a$ sobre el posible valor del $s_{max}$ no es mon\'otono como si sucede con otros par\'ametros (e.g. el efecto de los par\'ametros cu\'anticos $\omega $ y $\gamma$). Los resultados dan con la existencia de un mayor valor de esp\'in m\'aximo $s_{max}=1.6865$ cuando $a=0.3$. Conforme el ANR de Kerr se hace m\'as cercano al caso extremo ($a=M$), luego de que $s_{max}$ aumenta hasta que el par\'ametro $a=0.3$, el valor del $s_{max}$ disminuye inclusive m\'as que en el caso est\'atico de Schwarzschild.

 \begin{table}[H]
	\caption{\small{Valores del $s_{max}$ obtenido por el programa de Mathematica para una PP en el espacio-tiempo de un ANR modificado cu\'anticamente (con $G_0=1$ y $\omega=0.2$).}    \label{tabla4}}
	\centering {\small
		\begin{tabular}{cccc}\hline
			$\hspace{1cm}$ & $\hspace{1cm}$ & $s_{max}/M$ & $\hspace{1cm}$\\ \hline 
   $a/M$  &
		$\gamma=0.5$ & $\gamma=1$  & $\gamma=2$ \\ \hline
			$0.01$    & $ 1.7271$  &  $1.7720$ &  $1.8918$\\
			$0.05$    & $ 1.7368$  &  $1.7827$ &  $1.9064$\\
			$0.1$    & $ 1.7471$  &  $1.7942$ &  $1.9237$\\
			$0.2$    & $ 1.7607$  &  $1.8109$ &  $1.9555$\\
   		$0.3$    & $ 1.7639$  &  $1.8184$ &  $1.9872$\\
			$0.4$    & $ 1.7553$  &  $1.8161$ &  $2.0314$\\
   		$0.5$    & $ 1.7328$  &  $1.8047$ &  $2.1542$\\
			$0.6$    & $ 1.6940$  &  $1.7925$ &  No es AN\\
   		$0.7$    & $ 1.6396$  &  No es AN &  No es AN\\
     	$0.8$    & No es AN  &  No es AN &  No es AN\\
\hline
		\end{tabular}}
	\end{table}

 \begin{table}[H]
	\caption{\small{Valores del $s_{max}$ obtenido por el programa de Mathematica para una PP en el espacio-tiempo de un ANR modificado cu\'anticamente (con $G_0=1$ y $\omega=0.3$).}   \label{tabla5}}
	\centering {\small
		\begin{tabular}{cccc}\hline
			$\hspace{1cm}$ & $\hspace{1cm}$ & $s_{max}/M$ & $\hspace{1cm}$\\ \hline 
   $a/M$  &
		$\gamma=0.5$ & $\gamma=1$  & $\gamma=2$ \\ \hline
			$0.01$    & $ 1.7855$  &  $1.8834$ &  $2.2298$\\
			$0.05$    & $ 1.7966$  &  $1.8979$ &  $2.2685$\\
			$0.1$    & $ 1.8086$  &  $1.9149$ &  $2.3242$\\
			$0.2$    & $ 1.8263$  &  $1.9462$ &  $2.4833$\\
   		$0.3$    & $ 1.8352$  &  $1.9775$ &  $2.8192$\\
			$0.4$    & $ 1.8354$  &  $2.0229$ &  No es AN\\
   		$0.5$    & $ 1.8304$  &  $2.1701$ &  No es AN\\
			$0.6$    & $ 1.8523$  &  No es AN &  No es AN\\
   		$0.7$    & No es AN  &  No es AN &  No es AN\\
\hline
		\end{tabular}}
	\end{table}

 \begin{table}[H]
	\caption{\small{Valores del $s_{max}$ obtenido por el programa de Mathematica para una PP en el espacio-tiempo de un ANR modificado cu\'anticamente (con $G_0=1$ y $\omega=0.4$).} \label{tabla6}}
	\centering {\small
		\begin{tabular}{cccc}\hline
			$\hspace{1cm}$ & $\hspace{1cm}$ & $s_{max}/M$ & $\hspace{1cm}$\\ \hline 
   $a/M$  &
		$\gamma=0.5$ & $\gamma=1$  & $\gamma=2$ \\ \hline
			$0.01$    & $ 1.8724$  &  $2.0833$ &  No es AN\\
			$0.05$    & $ 1.8866$  &  $2.1106$ &  No es AN\\
			$0.1$    & $ 1.9032$  &  $2.1476$ &  No es AN\\
			$0.2$    & $ 1.9334$  &  $2.2427$ &  No es AN\\
   		$0.3$    & $ 1.9630$  &  $2.4212$ &  No es AN\\
			$0.4$    & $ 2.0076$  &  No es AN &  No es AN\\
   		$0.5$    & No es AN  &  No es AN &  No es AN\\
\hline
		\end{tabular}}
	\end{table}
Las tablas \ref{tabla2}, \ref{tabla3}, \ref{tabla4}, \ref{tabla5} y  \ref{tabla6} se muestran los valores del $s_{max}$ para diferentes valores posibles de los par\'ametros cu\'anticos $\omega $ y $\gamma$ de los ANs modificados cu\'anticamente. En estas se indican los valores de los par\'ametros $\omega $ y $\gamma$ en los cuales no se obtendr\'ia una soluci\'on de tipo AN. En la tabla \ref{tabla2} se muestra el valor del $s_{max}$ para diferentes valores posibles de los par\'ametros cu\'anticos $\omega $ y $\gamma$ del AN est\'atico modificado cu\'anticamente. En estos resultados se ve claramente que a mayores valores de los par\'ametros cu\'anticos $\omega $ y $\gamma$, el valor del $s_{max}$ ser\'a mayor. 

En las tablas \ref{tabla3}, \ref{tabla4}, \ref{tabla5} y  \ref{tabla6}, se muestran los resultados de los valores del $s_{max}$ para diferentes valores posibles de los par\'ametros cu\'anticos $\omega $ y $\gamma$ y del esp\'in $a$ del ANR modificado cu\'anticamente. Los resultados obtenidos en la tabla \ref{tabla3} con $\omega=0.1 $ y $\gamma=0.5$, $1$ y $2$, muestran que el mayor valor del $s_{max}$ a\'un se obtiene cuando $a=0.3$ como en el caso del ANR de Kerr. Sin embargo, en las tablas \ref{tabla4}, \ref{tabla5} y  \ref{tabla6} cuando se toma a $\omega=0.2$, $0.3$ y $0.4$ respectivamente, este comportamiento previo del $s_{max}$ se pierde. All\'i se muestra que a mayores valores de los par\'ametros cu\'anticos $\omega $ y $\gamma$, el valor del $s_{max}$ se hace mayor. De modo que para los ANR modificado cu\'anticamente con $\omega = 0.2$ y $\gamma = 2$ ya se obtiene un comportamiento mon\'otonamente creciente del $s_{max}$ con respecto al valor del \'espin $a$ del ANR. 

De igual manera, los par\'ametros cu\'anticos $\omega $ y $\gamma$ no deben ser muy grandes. Por ejemplo, como se ilustra en tabla \ref{tabla6}, con $\omega \geq 0.4$ y $\gamma \geq 2$ ya no se obtienen soluciones de tipo AN. En estos casos en que no es una soluci\'on de tipo AN, el programa de Mathematica da error.

De todas formas, como se muestra en las figuras \ref{fig:RBH3}  y \ref{fig:RBH4},  los efectos del aumento del esp\'in de los ANRs sobre los par\'ametros de la ISCO son mon\'otonamente decrecientes. All\'i se puede ver que tanto para el ANR de Kerr como en el modificado cu\'anticamente, los par\'ametros de la ISCO disminuyen en todos los casos, tanto para PP sin esp\'in ($s=0$), como para PP en \'orbitas esp\'in-alineadas ($s>0$) y esp\'in-antialineadas ($s<0$) con respecto al plano ecuatorial del ANR. Adicionalmente, en las figuras \ref{fig:RBH3}  y \ref{fig:RBH4}  se corrobora el hecho de que los par\'ametros de la ISCO del ANR modificado cu\'anticamente son menores que los del ANR de Kerr, sin importar los valores de $s$ o de $a$, como previamente se hab\'ia descrito en la referencia \cite{Larranaga2022}.

En general, los resultados muestran que el comportamiento de los par\'ametros de la ISCO de una PP con esp\'in  entre los ANRs de Kerr y los modificados cu\'anticamente es similar. Se obtienen siempre par\'ametros de la ISCO un poco menores en el caso del ANR modificado cu\'anticamente. Sin embargo, el efecto de $a$ sobre el $s_{max}$ si difiere entre estos ANRs debido a las correcciones cu\'anticas dadas por $\omega$ y $\gamma$. El cuaderno del programa Mathematica presentado en el repositorio \cite{gitrepo} fue una herramienta muy practica para evaluar estos detalles del comportamiento de los par\'ametros de la ISCO y del valor $s_{max}$, por lo que podr\'ia ser utilizado para seguir estudiando estos mismos fen\'omenos sobre un ANR modificado cu\'anticamente o tambi\'en sobre otras soluciones de tipo AN con las que se pueda adaptar la rutina.  

\section{Conclusiones}

En esta investigaci\'on se present\'o la deducci\'on de las ecuaciones de movimiento para una PP con esp\'in en el espacio-tiempo de un ANR modificado cu\'anticamente. El estudio se condicion\'o a una PP con esp\'in movi\'endose en el plano ecuatorial de \'orbitas esp\'in-alineadas y esp\'in-antialineada con el ANR y luego se defini\'o a el correspondiente potencial efectivo del sistema. Se identificaron a los par\'ametros de la ISCO y se describi\'o la determinaci\'on del valor de esp\'in m\'aximo $s_{max}$ f\'isicamente posible dado por una velocidad como de tiempo. Luego, se present\'o una implementaci\'on computacional a trav\'es de un cuaderno del programa Mathematica en el respositorio \cite{gitrepo}, con el cual, se calcular\'on el esp\'in m\'aximo de la PP y los par\'ametros de la ISCO para diversos valores del esp\'in $s$ de la PP y de los par\'ametros geom\'etricos de $a$, $\omega $ y $\gamma$ del ANR modificado cu\'anticamente.

Los resultados obtenidos con el cuaderno del programa Mathematica son satisfactorios y consistentes con los previamente obtenidos en las referencias \cite{Larranaga2022,Zhang2018,Jefremov2015}. Inicialmente, se se\~{n}alaron algunos de los valores de los par\'ametros $\omega $ y $\gamma$ en los cuales no se obtendr\'ia una soluci\'on de tipo AN. Luego, los resultados obtenidos para los par\'ametros de la ISCO tanto para los ANs cl\'asicos ($ \omega =0$), como para los ANs modificados cu\'anticamente ($ \omega \neq 0$) disminuyen con un aumento del esp\'in $s$ o de los par\'ametros cu\'anticos $ \omega $ y $\gamma$. Adem\'as, tambi\'en para mayores valores de los par\'ametros cu\'anticos $\omega $ y $\gamma$ se obtienen mayores valores del esp\'in m\'aximo $s_{max}$ permitido. Asimismo, se mostr\'o, que al aumentar el \'espin $a$ de los ANRs, los par\'ametros de la ISCO decrecen mon\'otonamente, por lo que estos son m\'as peque\~{n}os que los de los casos de los ANs est\'aticos. Consecuentemente, se confirma que los par\'ametros de la ISCO del ANR modificado cu\'anticamente son menores que los del ANR de Kerr, para cualesquiera valores de $s$ y $a$.   En cambio,  para el ANR de Kerr se calcul\'o el valor del $s_{max}$ en funci\'on de $a$ y se mostr\'o que no posee un cambio mon\'otono, sino que este posee un mayor valor del $s_{max}=1.6865$ cuando $a=0.3$. Tambi\'en se encontr\'o que para mayores valores de los par\'ametros cu\'anticos $\omega $ y $\gamma$, el valor del $s_{max}$ ser\'a mayor para la PP en los alrededores de un AN de Schwarzschild modificado cu\'anticamente. Por otro lado, para la PP en los alrededores de un ANR  modificado cu\'anticamente se hall\'o un comportamiento del $s_{max}$ que difiere de todos los casos particulares. En general, se encontr\'o que el valor del $s_{max}$ en el ANR  modificado cu\'anticamente no var\'ia mon\'otonamente con respecto a los par\'ametros geom\'etricos de $a$, $\omega $ y $\gamma$, como si sucede en los casos est\'aticos particulares, ni tampoco se obtiene un comportamiento con un pico m\'aximo para el valor de $s_{max}$ como se obtuvo para el AN de Kerr. Todos los resultados anteriormente compartidos sugieren que el comportamiento de los par\'ametros de la ISCO de una PP con esp\'in  entre los ANRs de Kerr y los modificados cu\'anticamente son similares y estos resultan ser siempre un poco menores en el caso del ANR modificado cu\'anticamente. No obstante, se prob\'o que el efecto de $a$ sobre el $s_{max}$ si difiere entre estos ANRs.

Finalmente, se demostr\'o que el cuaderno del programa Mathematica sirvi\'o como una herramienta \'util y pr\'actica para calcular y analizar el comportamiento de los par\'ametros de la ISCO y del valor del $s_{max}$ de una PP orbitando a un ANR modificado cu\'anticamente. Esta implementaci\'on computacional puede ser utilizada de manera libre, bien sea para seguir estudiando estos mismos fen\'omenos sobre ANRs modificados cu\'anticamente o tambi\'en para extender la investigaci\'on sobre otras soluciones de tipo AN.

\noindent\footnotesize{{\bf Financiaci\'on:} Este trabajo fue financiado parcialmente por la Vicerrector\'ia de Investigaci\'on y la Facultad de Ciencias de la Universidad Nacional de Colombia a trav\'es del Proyecto registrado en c\'odigo HERMES 57057.}\\
\footnotesize{{\bf Declaraci\'on de conflicto de inter\'es:} Los autores manifiestan no tener conflictos de inter\'es.}

\end{document}